\documentclass[preprint,superscriptaddress,preprintnumbers,showpacs]{elsart}
\usepackage{amssymb}
\usepackage{graphicx}
\usepackage{dcolumn}
\usepackage{bm}
\usepackage{amsmath}
\usepackage{subfigure}
\usepackage{overpic}

\journal{Physics Letter B. Accepted for publication.}

\begin{document}

\begin{frontmatter}

\title{A new relation between quark and lepton mixing matrices}

\author[1]{Nan Qin},
\author[1,2]{Bo-Qiang Ma\corauthref{*}} \corauth[*]{Corresponding author.}\ead{mabq@pku.edu.cn}
\address[1]{School of Physics and State Key Laboratory of Nuclear Physics and
Technology, Peking University, Beijing 100871, China}
\address[2]{Center for High Energy
Physics, Peking University, Beijing 100871, China}

\begin{abstract}
We propose a new relation between quark mixing matrix and lepton
mixing matrix. Since the parameters in the quark sector are well
determined, we employ them to describe the mixing of leptons.
Phenomenologically, we study the neutrino oscillation probabilities
for different channels, which can be measured precisely in
forthcoming reactor and accelerator experiments. As an example of
the applicability of our assumption, CP violation in the lepton
sector is also discussed. In the latest T2K experiment, the range of
the mixing angle $\theta_{13}$ is measured, and our prediction of
$\theta_{13}$ is compatible with their result.
\end{abstract}


\end{frontmatter}

The results of many neutrino oscillation experiments in the last
decade have demonstrated that there exist physics beyond standard
model in neutrino sector. It is commonly accepted that neutrinos are
massive and mixing~\cite{rev}. Neutrino oscillation is governed by
two mass square differences $\Delta m_{21}^2$, $\Delta m_{31}^2$ and
the lepton mixing matrix proposed by Pontecorvo, Maki, Nakawaga and
Sakata (PMNS)~\cite{pmns}
\begin{align}
U_{\rm PMNS}=\left(
  \begin{array}{ccc}
    U_{e1}    & U_{e2}    & U_{e3}    \\
    U_{\mu1}  & U_{\mu2}  & U_{\mu3}  \\
    U_{\tau1} & U_{\tau2} & U_{\tau3} \\
  \end{array}
\right).\label{pmns}
\end{align}
If neutrinos are of the Majorana type, there should be an additional
diagonal matrix with two Majorana phases $P={\rm
diag}(e^{i\alpha_1/2},e^{i\alpha_2/2},1)$ multiplied to
Eq.~(\ref{pmns}) from the right. But the two Majorana phases do not
affect neutrino oscillations, thus we do not include them in our
calculations.

Before more underlying theory of the origin of the mixing is found,
parametrizing the mixing matrix properly is helpful in understanding
the mixing pattern and analyzing experimental results. A commonly
used form is the standard parametrization proposed by Chau and Keung
(CK)~\cite{ck}
\begin{align}
U_{\rm CK}=\left(
\begin{array}{ccc}
c_{12}c_{13} & s_{12}c_{13} & s_{13}e^{-i\delta}           \\
-s_{12}c_{23}-c_{12}s_{23}s_{13}e^{i\delta} &
c_{12}c_{23}-s_{12}s_{23}s_{13}e^{i\delta}  & s_{23}c_{13} \\
s_{12}s_{23}-c_{12}c_{23}s_{13}e^{i\delta}  &
-c_{12}s_{23}-s_{12}c_{23}s_{13}e^{i\delta} & c_{23}c_{13}
\end{array}
\right),
\label{standard}
\end{align}
where $s_{ij}=\sin\theta_{ij}$ and $c_{ij}=\cos\theta_{ij}$
$(i,j=1,2,3)$ are the mixing angles, and $\delta$ is the
CP-violating phase. The current global fits of the neutrino mixing
parameters are given at the $1(3)\sigma$ level by~\cite{fit}
\begin{align}
\!\!\theta_{12}=34.4\pm1.0~(^{+3.2}_{-2.9})^{\circ},
~~\theta_{23}=42.8^{+4.7}_{-2.9}~(^{+10.7}_{-7.3})^{\circ},
~~\theta_{13}=5.6^{+3.0}_{-2.7}~(\leq12.5)^{\circ};\label{angles}
\end{align}
\begin{align}
\Delta m_{21}^2=7.59\pm0.20~(^{+0.61}_{-0.69})~\times10^{-5}{\rm eV}^2,\quad
\Delta m_{31}^2 =
    \left\lbrace
    \begin{matrix}
      -2.36 \pm 0.11 \,\left(\pm 0.37\right)
      \times 10^{-3}~{\rm eV}^2  \, ,
      \\[1mm]
      +2.46 \pm 0.12 \,\left(\pm{0.37}\right)
      \times 10^{-3}~{\rm eV}^2 \,.
    \end{matrix}
    \right.
\label{datas}
\end{align}
These results for angles are compatible with the tri-bimaximal (TB) matrix~\cite{tri}
\begin{align}
U_{\rm{TB}}=\left(
              \begin{array}{ccc}
              2/\sqrt{6}  & 1/\sqrt{3}  & 0          \\
              -1/\sqrt{6} & 1/\sqrt{3}  & 1/\sqrt{2} \\
              1/\sqrt{6}  & -1/\sqrt{3} & 1/\sqrt{2}
              \end{array}
            \right).
\label{tb}
\end{align}
Therefore, it is widely accepted that $U_{\rm TB}$ is a good
approximation to reality~\cite{tri-dev}.

In contrast to the large mixing in lepton sector, the observed
Cabibbo-Kobayashi-Maskawa~\cite{ckm}(CKM) quark mixing matrix
$V_{\rm CKM}$ is close to the unit matrix. Although it seems that
the mixing of quarks and leptons are unrelated with each other,
there indeed exist phenomenological relations between mixing angles
called quark-lepton complementarity (QLC)~\cite{qlc}, given by
\begin{align}
\theta_{12}^Q+\theta_{12}=\pi/4,\quad\theta_{23}^Q+\theta_{23}=\pi/4,\quad\theta_{13}^Q\sim\theta_{13}\sim0.
\end{align}
We pointed out in previous works~\cite{prev} that the unit matrix
pattern for quark mixing is connected with the bimaximal matrix
pattern~\cite{bi} for lepton mixing through QLC relations. But
present data imply that the tri-bimaximal matrix $U_{\rm TB}$ is
closer to reality than bimaximal matrix. Therefore, a natural idea
is to connect the unit matrix in quark sector with the tri-bimaximal
matrix in lepton sector. Based on this consideration, we propose
here a simple relation between the lepton and quark mixing matrices
\begin{align}
V^{\dagger}_{\rm CKM}U_{\rm PMNS}V_{\rm CKM}=U_{\rm TB},\label{ansatz}
\end{align}
under an Ansatz that $U_{\rm PMNS}$ becomes an exact tri-bimaximal
mixing $U_{\rm TB}$ in a limit $V_{\rm CKM}=1$~\cite{similar}. Such
a relation maybe comes from certain flavor symmetries, and the
corresponding symmetry breaking effects may induce the deviations of
the observed $V_{\rm CKM}$ and $U_{\rm PMNS}$ from the exact unit
matrix and tri-bimaximal matrix. By assuming Eq.~(\ref{ansatz}), we
can employ one set of parameters to describe both quark and lepton
mixing matrices, thus Eq.~(\ref{ansatz}) can be regarded as the
quark-lepton complementarity in matrix form.

Currently, the quark mixing matrix is well determined. For the
Wolfenstein parametrization~\cite{wolf}
\begin{align}
V_{\rm CKM}=\left( \begin{array}{ccc}
1-\frac{1}{2}\lambda^2&\lambda&A\lambda^3(\rho-i\eta)\\
-\lambda&1-\frac{1}{2}\lambda^2&A\lambda^2\\
A\lambda^3(1-\rho-i\eta)&-A\lambda^2&1
\end{array}\right)
+\mathcal {O}(\lambda^4),
\label{wolf}
\end{align}
the up-to-date fit gives~\cite{pdg}
\begin{align}
&\lambda=0.2253\pm0.0007\;,
\quad\quad A=0.808_{-0.015}^{+0.022}\;,\nonumber\\
&\rho(1-\lambda^2/2+\ldots)=0.132_{-0.014}^{+0.022}\;, \quad
\eta(1-\lambda^2/2+\ldots)=0.341\pm0.013\;.\label{wolfdata}
\end{align}

By inserting Eq.~(\ref{wolf}) into Eq.~(\ref{ansatz}), one can
easily get the PMNS matrix in terms of $\lambda,~A,~\rho$ and
$\eta$, given by\footnote[1]{For simplicity, we present expressions
to the second order of $\lambda$ here, but all the results below
result from $U_{\rm PMNS}$ to $\mathcal{O}(\lambda^3)$.}
\begin{align}
U_{\rm PMNS}&=V_{\rm CKM}U_{\rm TB}V_{\rm CKM}^{\dagger}\nonumber\\
&=U_{\rm TB}+\lambda\left(
\begin{array}{ccc}
 \frac{1}{\sqrt{3}}-\frac{1}{\sqrt{6}} & -\frac{-1+\sqrt{2}}{\sqrt{3}} & \frac{1}{\sqrt{2}} \\
 -\frac{-1+\sqrt{2}}{\sqrt{3}} & -\frac{1}{\sqrt{3}}+\frac{1}{\sqrt{6}} & 0 \\
 -\frac{1}{\sqrt{3}} & -\frac{1}{\sqrt{6}} & 0
\end{array}\right)\nonumber\\
&+\lambda^2\left(\begin{array}{ccc}
 -\frac{-1+\sqrt{2}}{\sqrt{3}} & -\frac{1}{\sqrt{3}}+\frac{1}{\sqrt{6}} & -\frac{A}{\sqrt{3}} \\
 \frac{A}{\sqrt{6}}+\frac{1}{\sqrt{6}}-\frac{1}{\sqrt{3}} & \left(\frac{1}{\sqrt{2}}-\frac{1}{\sqrt{3}}\right) A+\frac{-1+\sqrt{2}}{\sqrt{3}} & \left(\frac{1}{\sqrt{2}}-\frac{1}{\sqrt{3}}\right) A-\frac{1}{2 \sqrt{2}} \\
 \frac{2 A-1}{2 \sqrt{6}} & \frac{1}{6} \left(-2 \sqrt{3} A+3 \sqrt{2} A+\sqrt{3}\right) & \left(-\frac{1}{\sqrt{2}}+\frac{1}{\sqrt{3}}\right) A
\end{array}
\right)
+\mathcal{O}(\lambda^3).\label{parametrization}
\end{align}
Clearly, we describe the deviations of $V_{\rm CKM}$ from the unit
matrix and $U_{\rm PMNS}$ from the tri-bimaximal matrix with the
same set of parameters. In other words, relation
Eq.~(\ref{ansatz}) provides a unified way of parametrizing both
quark and lepton mixing matrices.

The lepton mixing angles are given by
\begin{align}
&\sin^2{\theta_{13}}=\frac{\lambda^2}{2}-\sqrt{\frac{2}{3}}A\lambda^3+\mathcal{O}(\lambda^4),\nonumber\\
&\sin^2{\theta_{23}}=\frac{1}{2}-\frac{1}{12}\left(4(\sqrt{6}-3)A
+3\right)\lambda^2+\frac{A}{2\sqrt{3}}\left(2\rho+\sqrt{2}-2\right)\lambda^3+\mathcal{O}(\lambda^4),\nonumber\\
&\sin^2{\theta_{12}}=\frac{1}{3}-\frac{2}{3}(\sqrt{2}-1)\lambda+\frac{1}{6}(3-2\sqrt{2})\lambda^2
+\frac{1}{9}\left(2A(\sqrt{6}-3\rho)+9\sqrt{2}-12\right)\lambda^3+\mathcal{O}(\lambda^4).\label{sin}
\end{align}
With the central values in Eq.~(\ref{wolfdata}), we get numerically
\begin{align}
&\theta_{12}\cong31.80^{\circ},\quad\theta_{23}\cong44.66^{\circ},\quad\theta_{13}\cong7.67^{\circ},
\label{prediction}
\end{align}
which are compatible with the fit results in Eq.~(\ref{angles}). Comparing with the exact tri-bimaximal mixing angles
\begin{align}
\theta_{12}=\arcsin{\frac{1}{\sqrt{3}}},\quad\theta_{23}=\pi/4,\quad\theta_{13}=0,
\end{align}
our result for $\theta_{12}$ is more closer to the prediction of QLC since the $1\sigma$ range for the corresponding quark mixing angle reads~\cite{fitter} $\theta_{12}^Q=13.03(\pm0.05)^{\circ}$. Denoting $\epsilon_{ij}$ the deviations from the exact tri-bimaximal angles, we have
\begin{align}
\epsilon_{12}\equiv\theta_{12}-\arcsin{\frac{1}{\sqrt{3}}}\cong-3.46^{\circ},\quad\epsilon_{23}\equiv\theta_{23}-\frac{\pi}{4}\cong-0.34^{\circ},\quad\epsilon_{13}\equiv\theta_{13}\cong7.67^{\circ}.
\end{align}
As it shows, a relatively large $\theta_{13}$ is predicted from our
assumption, and such a prediction can be tested precisely in future
reactor and accelerator experiments.

An important property of the relation in Eq.~(\ref{ansatz}) is that
different phase conventions of $V_{\rm CKM}$ give different
predictions on lepton mixing angles. We give a brief argument here.
If $V_{\rm CKM}$ is rephased by taking
\begin{align}
V'_{\rm CKM}=\Psi_1V_{\rm CKM}\Psi_2^{\dagger},
\end{align}
in which $\Psi_1\equiv{\rm diag}(e^{iu},e^{ic},e^{it})$ and
$\Psi_2\equiv{\rm diag}(e^{id},e^{is},e^{ib})$ consist of arbitrary
phases and can be absorbed into the redefinition of quark phases,
Eq.~(\ref{ansatz}) turns into
\begin{align}
U=V'_{\rm CKM}U_{\rm TB}{V_{\rm CKM}'}^{\dagger}=\Psi_1V_{\rm CKM}\Psi_2^{\dagger}U_{\rm TB}\Psi_2V_{\rm CKM}^{\dagger}\Psi_1^{\dagger}.\label{dep}
\end{align}
Because $V_{\rm CKM}$ ($V_{\rm CKM}^{\dagger}$) does not commute
with $\Psi_2^{\dagger}$ ($\Psi_2$), it is generally impossible to
absorb the phases in $\Psi_2$ ($\Psi_2^{\dagger}$) into the
redefinition of lepton phases. As a result, the magnitudes of
elements in $U_{\rm PMNS}$ and consequently lepton mixing angles
depend on the rephasing matrix $\Psi_2$ ($\Psi_2^{\dagger}$), which
does not bring any difference in $V_{\rm CKM}$. Therefore,
predictions on lepton mixing resulting from Eq.~(\ref{ansatz}) will
be changed if we change the parametrization of quark mixing matrix.
Similar arguments can be applied for assumptions discussed in
Ref.~\cite{similar}. For generality, detailed discussion concerning
the behavior of mixing angles on phases in $\Psi_2^{\dagger}$
($\Psi_2$) is given in the Appendix of this letter. As an instance,
another parametrization~\cite{qinnan} for $V_{\rm CKM}$ is also
employed to determine leptonic mixing angles in the Appendix.

We now turn to the application of the assumption Eq.~(\ref{ansatz})
to neutrino oscillations. Let us denote
$P_{\alpha\beta}=P(\nu_{\alpha}\rightarrow\nu_{\beta})$ the
probability of transition from a neutrino flavor $\alpha$ to a
neutrino flavor $\beta$. Similar to the discussion in
Ref.~\cite{probability}, the probability can be found as
$P_{\alpha\beta}=|S_{\beta\alpha}(t,t_0)|^2$, in which $S(t,t_0)$ is
the evolution matrix such that
\begin{align}
|\nu(t)\rangle=S(t,t_0)|\nu(t_0)\rangle,~~~~~~~~~~~S(t_0,t_0)=1.
\end{align}
For simplicity, we neglect the effects due to interactions between
neutrinos and matters in which the neutrino beam propagates and only
deal with oscillation probabilities in vacuum, the evolution matrix
can be given by
\begin{align}
S_{\beta\alpha}(t,t_0)=\sum_{i=1}^3(U_{\alpha i})^*U_{\beta i}e^{-iE_iL},~~~~~~~~\alpha,~\beta=e,\mu,\tau,
\label{evolution}
\end{align}
where $L=t-t_0$ is the length of the baseline in neutrino experiment and  $E_i$ are the eigenvalue of the effective hamiltonian
\begin{align}
H\simeq\frac{1}{2E}U{\rm diag}(0,\Delta m_{21}^2,\Delta m_{31}^2)U^{\dagger},
\end{align}
in which $E$ is the average energy of the neutrino beam.

Inspecting the values of the mass square difference in
Eq.~(\ref{datas}), one can find $\Delta m^2_{21}$ is much less than
$\Delta m^2_{31}$, thus can be neglected to a good precision. The
calculation of $P_{\alpha\beta}$ is now straightforward, such that
by combining Eq.~(\ref{ansatz}), Eq.~(\ref{wolf}) and
Eq.~(\ref{evolution}), we get the evolution matrix $S_{\beta\alpha}$
and consequently the oscillation probability $P_{\alpha\beta}$.
Expanded to $\lambda^3$, oscillation probabilities can be expressed
in a matrix form as
\begin{align}
P=\left(
\begin{array}{ccc}
1-2\lambda^2\Delta+4\sqrt{\frac{2}{3}}A\lambda^3\Delta & \lambda^2\Delta-2\sqrt{\frac{2}{3}}A\lambda^3\Delta & \lambda^2\Delta-2\sqrt{\frac{2}{3}}A\lambda^3\Delta \\
\lambda^2\Delta-2\sqrt{\frac{2}{3}}A\lambda^3\Delta & 1-\Delta & \Delta-\lambda^2\Delta+2\sqrt{\frac{2}{3}}A\lambda^3\Delta \\
\lambda^2\Delta-2\sqrt{\frac{2}{3}}A\lambda^3\Delta & \Delta-\lambda^2\Delta+2\sqrt{\frac{2}{3}}A\lambda^3\Delta & 1-\Delta
\end{array}
\right)+\mathcal{O}(\lambda^4),\label{pro}
\end{align}
where we have defined $\Delta\equiv \sin^2{(\frac{L\Delta m_{31}^2}{4E})}$.

Let us have a first look at the oscillation probability matrix in
Eq.~(\ref{pro}). Apparently, it exhibits a hierarchical structure
among different channels of neutrino oscillation. The diagonal
elements $P_{ee}$, $P_{\mu\mu}$ and $P_{\tau\tau}$ which measure the
disappearance probabilities for $\nu_e$, $\nu_{\mu}$ and
$\nu_{\tau}$ beams are of $\mathcal{O}(1)$. It is also not difficult
to understand that the $\nu_{\mu}\leftrightarrow\nu_{\tau}$
probabilities $P_{\mu\tau}$ and $P_{\tau\mu}$ are of
$\mathcal{O}(1)$, since $\nu_{\mu}$ and $\nu_{\tau}$ are maximally
mixing implied by data. Other terms measuring probabilities of
$\nu_{e}\leftrightarrow\nu_{\mu}$ and
$\nu_{e}\leftrightarrow\nu_{\tau}$ are of $\mathcal{O}(\lambda^2)$.
Interestingly, there are no terms proportional to $\lambda$, that is
because such terms are suppressed by $\Delta m_{21}^2$, which we
neglect here.

To get the anti-neutrino oscillation probabilities
$P_{\bar{\alpha}\bar{\beta}}$, one needs to take the replacement
$U_{\rm PMNS}\rightarrow U_{\rm PMNS}^*$, which, in our case, is
equivalent to reverse the sign of $\eta$, i.e.
\begin{align}
P_{\bar{\alpha}\bar{\beta}}=P_{\alpha\beta}(\eta\rightarrow-\eta).
\end{align}
However, the CP-violating parameter $\eta$ is missing in Eq.~(\ref{pro}), meaning that to this approximation, we have
\begin{align}
P_{\bar{\alpha}\bar{\beta}}=P_{\alpha\beta}.
\end{align}
Therefore, CP symmetry is preserved, and resulting from the CPT
theorem, time reversal symmetry is also preserved, i.e.,
$P_{\alpha\beta}=P_{\beta\alpha}$. Consistently, Eq.~(\ref{pro}) is
a symmetric matrix.

Defining the asymmetries in neutrino oscillations as
\begin{align}
A_{\alpha\beta}^{\rm CP}=P_{\alpha\beta}-P_{\bar{\alpha}\bar{\beta}},
\end{align}
the probabilities in Eq.~(\ref{pro}) imply $A_{\alpha\beta}^{\rm
CP}=0$. The reason for the vanishing of $A_{\alpha\beta}^{\rm CP}$
is that we neglect the contribution of the smaller mass square
difference $\Delta m_{21}^2$. Now taking this into account, we get
nonzero values for CP asymmetries as
\begin{align}
A^{\rm CP}=\left(\begin{array}{ccc}
0&a&-a\\
-a&0&a\\
a&-a&0\end{array}\right),\label{CP}
\end{align}
in which
\begin{align}
a=\frac{2}{9}(2\sqrt{3}-3)A\eta\lambda^3\left(\sin{(\frac{L\Delta m_{31}^2}{2E})}-\sin{(\frac{L(\Delta m_{31}^2-\Delta m_{21}^2)}{2E})}-\sin{(\frac{L\Delta m_{21}^2}{2E})}\right)+\mathcal{O}(\lambda^4).\label{cpv}
\end{align}
The structure of asymmetries in Eq.~(\ref{CP}) comes from the
unitarity of PMNS matrix and the conservation of probability. One
also has~\cite{petcov}
\begin{align}
A_{\mu e}^{\rm CP}=-A_{\tau e}^{\rm CP}=A_{\tau\mu}^{\rm CP}=4J_{\rm CP}\left(\sin{(\frac{L\Delta m_{13}^2}{2E})}+\sin{(\frac{L\Delta m_{32}^2}{2E})}+\sin{(\frac{L\Delta m_{21}^2}{2E})}\right),\label{pet}
\end{align}
in which
\begin{align}
J_{\rm CP}\equiv{\rm{Im}}(U_{\mu3}U_{e2}U_{e3}^\ast U_{\mu2}^\ast)
\end{align}
is the rephasing invariant~\cite{jarlskog}. Combining Eq.~(\ref{CP}), Eq.~(\ref{cpv}) and Eq.~(\ref{pet}), we get
\begin{align}
J_{\rm CP}=-\frac{1}{18}(2\sqrt{3}-3)A\eta\lambda^3+\mathcal{O}(\lambda^4).\label{jcp}
\end{align}
Compared with the results in Ref.~\cite{similar}, where they arrive
at $J_{\rm CP}\sim\mathcal{O}(\lambda)$ by assuming $V_{\rm
CKM}U_{\rm PMNS}=U_{\rm TB}$ or $V_{\rm CKM}^{\dagger}U_{\rm
PMNS}=U_{\rm TB}$, prediction of $J_{\rm CP}$ here is quite smaller.
This is because that the exact $U_{\rm TB}$ implies $J_{\rm CP}=0$,
thus $J_{\rm CP}$ only depends on the deviation of $U_{\rm PMNS}$
from $U_{\rm TB}$ (which we denote by $D$ below). Since $V_{\rm
CKM}$ is close to the unit matrix, it can be regarded as the
measurement of $D$. Then it is not difficult to understand the
smallness of $J_{\rm CP}$ in Eq.~(\ref{jcp}) as one has
\begin{align}
D\sim\mathcal{O}(V_{\rm CKM})\quad{\rm and}\quad J_{CP}\sim\mathcal{O}(\lambda)
\end{align}
in Ref.~\cite{similar}, while
\begin{align}
D\sim\mathcal{O}(V_{\rm CKM}^2)\quad{\rm and}\quad J_{CP}\sim\mathcal{O}(\lambda^3)
\end{align}
in this letter.

We emphasize that in the very recent T2K experiment~\cite{t2k},
observations of $\nu_{\mu}\rightarrow\nu_e$ oscillation indicate
that at $90\%$ C.L., the data are consistent with
\begin{align}
0.03(0.04)<\sin^2{2\theta_{13}}<0.28(0.34)\label{theta13}
\end{align}
for $\delta=0$ and normal (inverted) hierarchy. Such a result implies an apparent deviation from $U_{\rm TB}$ and is important in the establishment of lepton mixing pattern, in which new symmetries among leptons may hide. If $\theta_{13}$ is really large, the test of CP violation in the lepton sector is possible for future neutrino experiments since $\delta$ is always multiplied by $\theta_{13}$ in $U_{\rm PMNS}$. With Eq.~(\ref{theta13}), straightforward calculations give
\begin{align}
4.99^{\circ}(5.77^{\circ})<\theta_{13}<15.97^{\circ}(17.83^{\circ})
\end{align}
at $90\%$ C.L., for $\delta=0$ and normal (inverted) hierarchy. We find that our prediction, i.e., $\theta_{13}\cong7.67$ is compatible with the T2K result. Thus the relation Eq.~(\ref{ansatz}) may serve as a good description of the deviation of $U_{\rm PMNS}$ from $U_{\rm TB}$.

In summary, we propose a new relation between quark mixing matrix
$V_{\rm CKM}$ and lepton mixing matrix $U_{\rm PMNS}$ given by
Eq.~(\ref{ansatz}) as the quark-lepton complementarity in matrix
form. Based on this relation, we parametrize $U_{\rm PMNS}$ with the
quark mixing parameters in Eq.~(\ref{parametrization}), and
determine the deviations of mixing angles from the exact
tri-bimaximal angles. Especially, our prediction of the mixing angle
$\theta_{13}$ agrees with the latest T2K result. For neutrino
oscillations, we derive oscillation probabilities for different
channels given by Eq.~(\ref{pro}). As we can see, to a good
precision, the expressions for each probability are quite simple.
Furthermore, the CP violation in neutrino flavor transitions are
discussed.

This work is partially supported by National Natural Science
Foundation of China (Grants No.~11021092, No.~10975003,
No.~11035003) and by the Key Grant Project of Chinese Ministry of
Education (Grant No.~305001).

\appendix

{\bf {Appendix : Dependence of lepton mixing angles on the phase convention of CKM matrix}}

With discussions concerning Eq.~(\ref{dep}), we have pointed out that the predicted lepton mixing angles are dependent on the phase convention of $V_{\rm CKM}$, i.e., $\Psi_2$ ($\Psi_2^{\dagger}$) in our notation. By substituting the Wolfenstein matrix Eq.~(\ref{wolf}) into Eq.~(\ref{dep}), we arrive at the general expression of $U_{\rm PMNS}$, which includes explicitly the phases in $\Psi_1$ and $\Psi_2$. Since phases in $\Psi_1$ ($\Psi_1^{\dagger}$) can be absorbed into the redefinition of lepton fields, the physical mixing angles are thus dependent only on phase parameters $d$, $s$ and $b$. To a good accuracy, mixing angles are determined by
\begin{align}
&\sin^2{\theta_{12}}=\frac{1}{3}-\frac{2}{3} \lambda  \left(\sqrt{2} \cos {\alpha} -\cos {\alpha}\right)+\frac{1}{6} \lambda ^2 \left(2 \sqrt{2} \cos {2 \alpha }-4 \sqrt{2}+3\right)+\mathcal{O}(\lambda^3),\nonumber\\
&\sin^2{\theta_{23}}=\frac{1}{2}+\lambda ^2 \left(-\sqrt{\frac{2}{3}} A \cos {\beta }+A \cos {\beta }-\frac{1}{4}\right)+\mathcal{O}(\lambda^3),\nonumber\\
&\sin^2{\theta_{13}}=\frac{\lambda ^2}{2}-\sqrt{\frac{2}{3}} A \lambda ^3 \cos {\gamma }+\mathcal{O}(\lambda^4),\nonumber
\end{align}
where we define $\alpha\equiv d-s$, $\beta\equiv b-s$ and $\gamma\equiv b+d-2s$, and the last equation is to $\mathcal{O}(\lambda^3)$ because of the smallness of $\theta_{13}$. It is easy to see that the sensitivities to phases of $\sin^2{\theta_{12}}$, $\sin^2{\theta_{23}}$ and $\sin^2{\theta_{13}}$ are of $\mathcal{O}(\lambda)$, $\mathcal{O}(\lambda^2)$ and $\mathcal{O}(\lambda^3)$ respectively. With the best fit values in Eq.~(\ref{wolfdata}), the dependence of each mixing angle is illustrated in Fig.~\ref{ss} and Fig.~\ref{ss13}, which show that, the prediction of $\theta_{12}$ is strongly dependent on the phase, such that some areas of $\alpha$ are excluded by data and, the result in Eq.~(\ref{prediction}) can be improved by choose a particular value for $\alpha$. However, the dependence of the predicted $\theta_{23}$ and $\theta_{13}$ on phases is negligibly small compared with current data.
\renewcommand\thefigure{\arabic{figure}}
\begin{figure}[!htb]
\centering
\subfigure{
\includegraphics[width=0.45\textwidth]{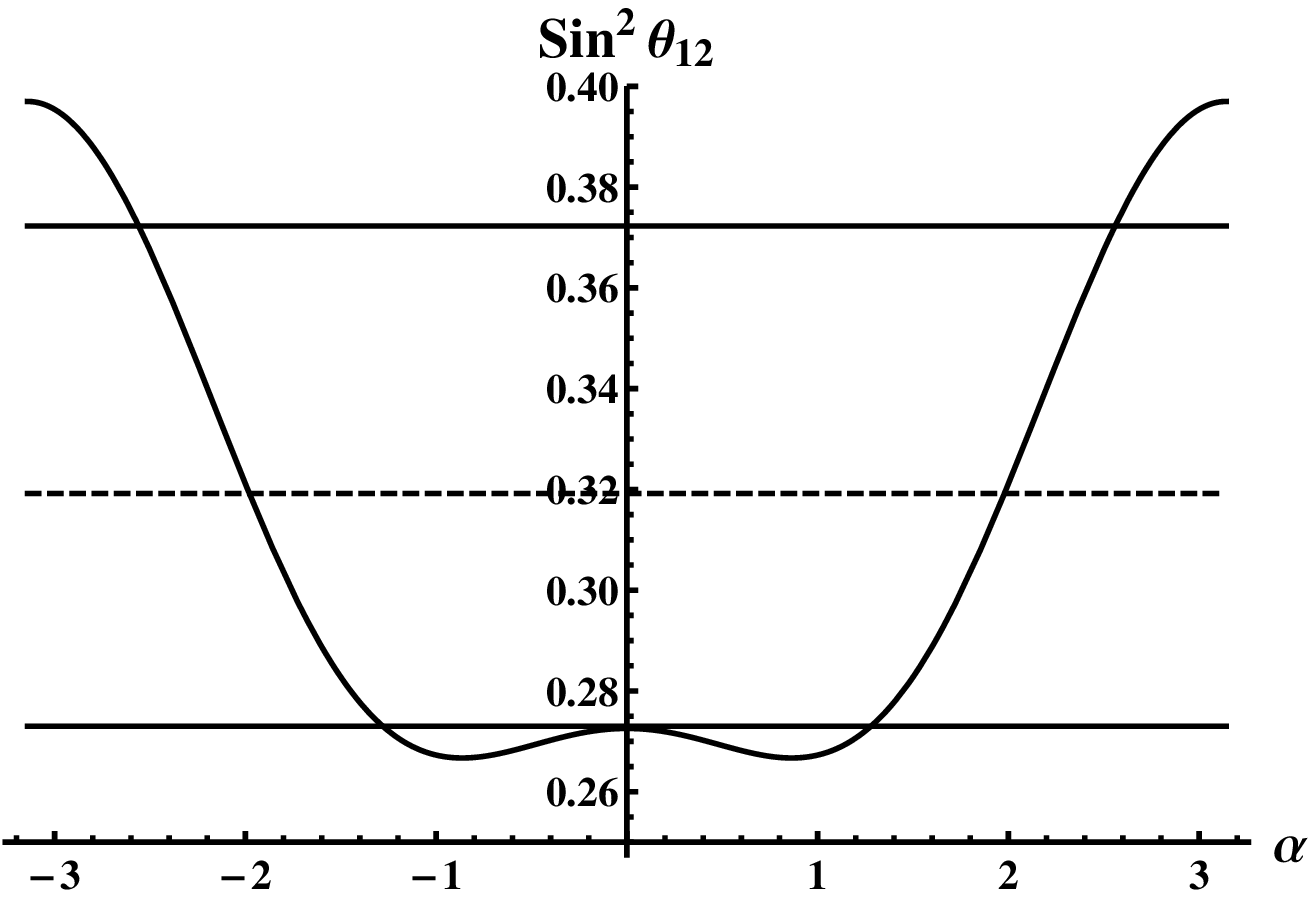}}
\hspace{0.5cm}
\subfigure{
\includegraphics[width=0.45\textwidth]{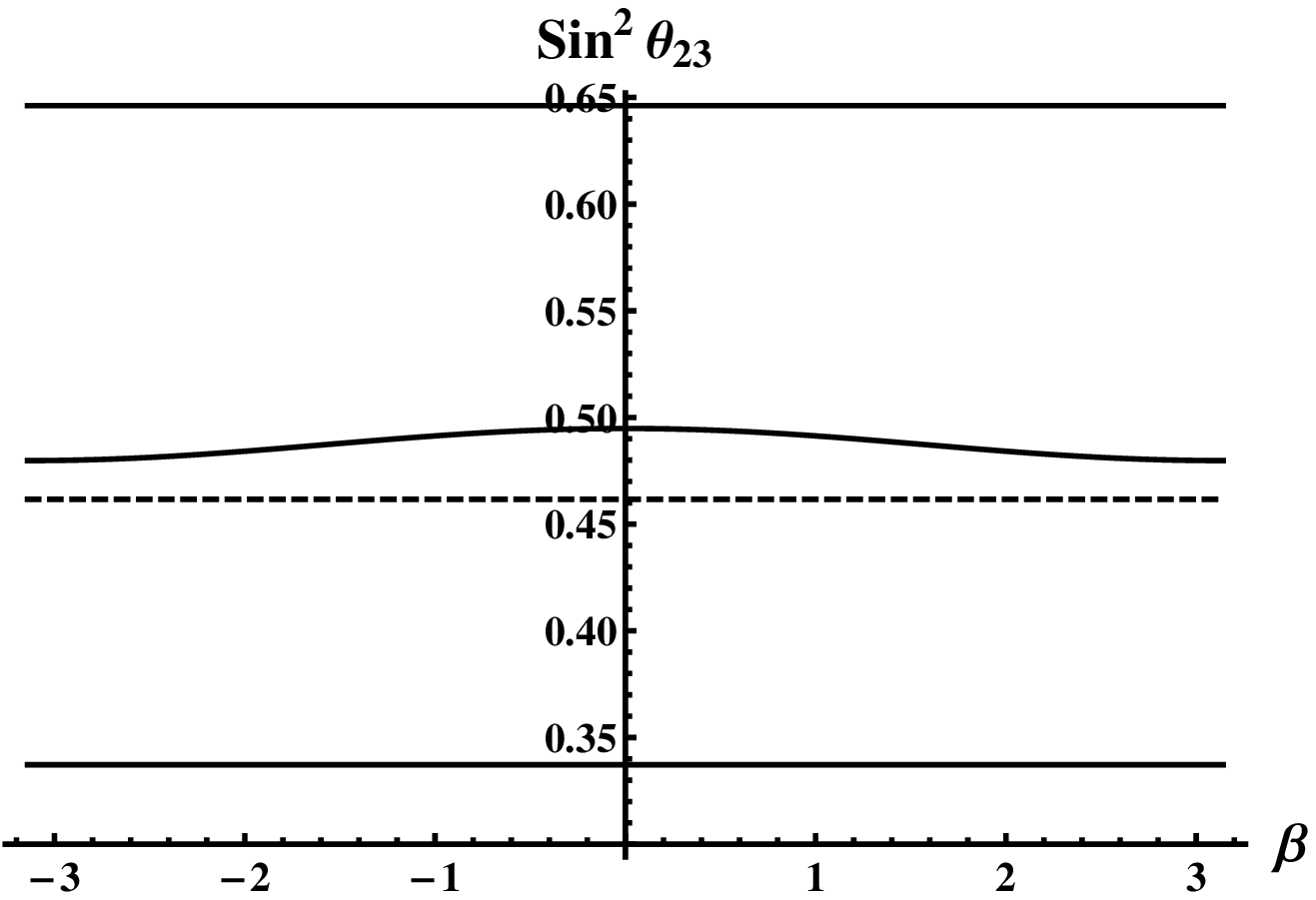}}
\caption{Behavior of $\sin^2{\theta_{12}}$ versus $\alpha\equiv d-s$
(left) and $\sin^2{\theta_{23}}$ versus $\beta\equiv b-s$ (right),
both from $-\pi$ to $\pi$. The solid horizontal lines denote the
$3\sigma$ ranges calculated from Eq.~(\ref{angles}), while the
dashed lines denote the best fit values.\label{ss}}
\end{figure}
\renewcommand\thefigure{\arabic{figure}}
\begin{figure}[!htb]
\centering
\includegraphics[width=0.5\textwidth]{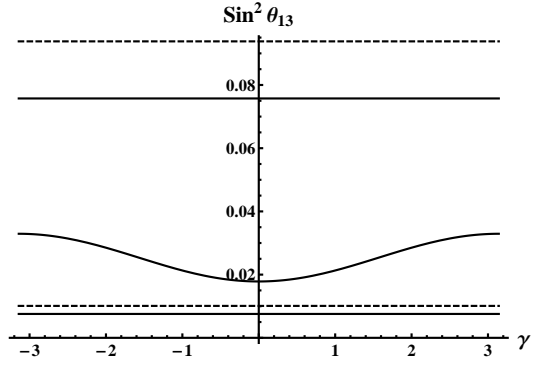}
\caption{Behavior of $\sin^2{\theta_{13}}$ versus $\gamma\equiv
b+d-2s$ from $-\pi$ to $\pi$. The solid (dashed) horizontal lines
denote $90\%$ C.L. ranges of the newest T2K result with normal
(invertd) hierarchy.}\label{ss13}
\end{figure}

In order to demonstrate that our predictions of $U_{\rm PMNS}$ and
mixing angles depend on the parametrization of $V_{\rm CKM}$, we
here employ another Wolfenstein-like parametrization~\cite{qinnan},
given by
\begin{align}
V_{\rm CKM}=\left(\begin{array}{ccc}
1-\frac{\lambda^2}{2}&\lambda&e^{-i\delta}h\lambda^3\\
-\lambda&1-\frac{\lambda^2}{2}&(f+e^{-i\delta}h)\lambda^2\\
f\lambda^3&-(f+e^{i\delta}h)\lambda^2&1
\end{array}\right)+\mathcal{O}(\lambda),\nonumber\label{newwolf}
\end{align}
to deduce the form of $U_{\rm PMNS}$. The ranges for the parameters
are
\begin{align}
\lambda=0.2253\pm{0.0007},\quad h=0.303^{+0.014}_{-0.010},\quad
f=0.754^{+0.022}_{-0.018},\quad
\delta^Q={90.97^{\circ}}_{-4.44^{\circ}}^{+2.77^{\circ}}.\nonumber
\end{align}
Further detailed analysis and discussions concerning the
relationship of this form of parametrization with others are given
in Ref.~\cite{Ahn:2011it}. By substituting this new Wolfenstein-like
parametrization into Eq.~(\ref{ansatz}), we get the expression for
$U_{\rm PMNS}$ to $\mathcal{O}(\lambda^2)$ as
\begin{small}
\begin{align}
&U_{\rm PMNS}=V_{\rm CKM}U_{\rm TB}V_{\rm CKM}^{\dagger} =U_{\rm
TB}+\lambda\left(
\begin{array}{ccc}
 \frac{1}{\sqrt{3}}-\frac{1}{\sqrt{6}} & -\frac{-1+\sqrt{2}}{\sqrt{3}} & \frac{1}{\sqrt{2}} \\
 -\frac{-1+\sqrt{2}}{\sqrt{3}} & -\frac{1}{\sqrt{3}}+\frac{1}{\sqrt{6}} & 0 \\
 -\frac{1}{\sqrt{3}} & -\frac{1}{\sqrt{6}} & 0
\end{array}
\right)\nonumber\\
&+\lambda^2\left(
\begin{array}{ccc}
 -\sqrt{\frac{2}{3}}+\frac{1}{\sqrt{3}} & -\frac{1}{\sqrt{3}}+\frac{1}{\sqrt{6}} & -\frac{f}{\sqrt{3}}-\frac{e^{-i \delta }}{\sqrt{3}} \\
 \frac{f}{\sqrt{6}}+\frac{e^{-i \delta } h}{\sqrt{6}}+\frac{1}{\sqrt{6}}-\frac{1}{\sqrt{3}} & -\frac{f}{\sqrt{3}}+\frac{f}{\sqrt{2}}-\frac{e^{-i
\delta } h}{\sqrt{3}}+\frac{e^{i \delta }
h}{\sqrt{2}}-\frac{1}{\sqrt{3}}+\sqrt{\frac{2}{3}} &
-\frac{f}{\sqrt{3}}+\frac{f}{\sqrt{2}}-\frac{e^{-i
\delta }}{\sqrt{3}}+\frac{e^{-i \delta } h}{\sqrt{2}}-\frac{1}{2 \sqrt{2}} \\
 \frac{f}{\sqrt{6}}+\frac{e^{i \delta }}{\sqrt{6}}-\frac{1}{2 \sqrt{6}} & -\frac{f}{\sqrt{3}}+\frac{f}{\sqrt{2}}-\frac{e^{i \delta }}{\sqrt{3}}+\frac{e^{i
\delta } h}{\sqrt{2}}+\frac{1}{2 \sqrt{3}} &
\frac{f}{\sqrt{3}}-\frac{f}{\sqrt{2}}+\frac{e^{-i \delta
}}{\sqrt{3}}-\frac{e^{i \delta }}{\sqrt{2}}
\end{array}
\right).\nonumber
\end{align}
\end{small}
Comparing with Eq.~(\ref{parametrization}), one can easily find that
differences begin to appear in terms of $\mathcal{O}(\lambda^2)$. To
$\mathcal{O}(\lambda^3)$, mixing angles are given by
\begin{small}
\begin{align}
&\sin^2{\theta_{13}}=\frac{\lambda^2}{2}-\sqrt{\frac{2}{3}}(f+\cos{\delta})\lambda^3,\nonumber\\
&\sin^2{\theta_{23}}=\frac{1}{2}-\frac{1}{12}\left(4(\sqrt{6}-3)f+4(\sqrt{6}-3h)\cos{\delta}+3\right)\lambda^2 +\frac{1}{2\sqrt{3}}\left((\sqrt{2}-2)f+\sqrt{2}\cos{\delta}\right)\lambda^3,\nonumber\\
&\sin^2{\theta_{12}}=\frac{1}{3}-\frac{2}{3}(\sqrt{2}-1)\lambda+\frac{1}{6}(3-2\sqrt{2})\lambda^2
+\frac{1}{9}\left(2\sqrt{6}f+(3(\sqrt{6}-2)h-\sqrt{6})\cos{\delta}+9\sqrt{2}-12\right)\lambda^3,\nonumber
\end{align}
\end{small}
in which $\sin^2{\theta_{13}}$, $\sin^2{\theta_{12}}$ differ from
Eq.~(\ref{sin}) only in terms proportional to $\lambda^3$ and the
difference of $\sin^2{\theta_{23}}$ is of $\mathcal{O}(\lambda^2)$.
Therefore, the numerical results for mixing angles are very close to
Eq.~(\ref{prediction}), as straightforward calculation gives
\begin{align}
\theta_{12}\cong31.84^{\circ},\quad\theta_{23}\cong44.61^{\circ},\quad\theta_{13}\cong7.82^{\circ}.\nonumber
\end{align}

%\vspace{14cm}

\end{document}